\documentclass[13pt]{siamltex}
\usepackage{graphics,epsfig,graphicx,amssymb,latexsym}
\usepackage{fancyhdr}

\begin{document}


\title{Interactive visualization of higher dimensional data in a multiview environment} 

\author{Stanimire Tomov\and 
        Michael McGuigan\thanks{Information Technology Division, 
                         Brookhaven National Laboratory, 
                         Bldg. 515, Upton, NY 11973\newline
                       ~tomov@bnl.gov, ~mcguigan@bnl.gov}
}

\pagestyle{myheadings}
\thispagestyle{plain}

\markboth{Stanimire Tomov and Michael McGuigan}{INTERACTIVE VISUALIZATION IN A 
MULTIVIEW ENVIRONMENT}
\date{February 20, 2003}

\maketitle

\begin{center}
\today{}
\end{center}


\begin{abstract} 
We develop multiple view visualization of higher 
dimensional data.
Our work was chiefly motivated by the need to extract insight 
from four dimensional Quantum Chromodynamic (QCD) data. We develop
visualization where multiple views, generally views of 3D projections or
slices of a higher dimensional data, 
are tightly coupled not only by their specific order
but also by a view synchronizing interaction style, and an internally 
defined interaction language.
The tight coupling of the different views 
allows a fast and well-coordinated exploration of 
the data. In particular, the visualization allowed us to easily make
consistency checks of the 4D QCD data and to infer the correctness of
particle properties calculations. The software developed was also 
successfully applied in material studies, in particular studies of 
meteorite properties.
Our implementation uses the VTK API. To handle a large number of views 
(slices/projections) and to still maintain good resolution, we use 
{\sl IBM T221} display ($3840$ X $2400$ pixels).
\end{abstract}


\begin{keywords}
Multiple view, interactive visualization, Quantum Chromodynamic,
exploratory visualization, VTK, {\sl IBM T221} display.
\end{keywords}


\section{Introduction}\label{introduction}


Our goal is to visually analyze and extract insight from higher dimensional
data. Also, in order to capture important details in the data, we seek
to examine the full data, or at least as much as possible.
To this end we employ (1) interactive, (2) tightly coupled, (3) multiview 
visualization to a (4) high-resolution flat panel. 

One of our applications is visualization of Quantum Chromodynamic (QCD) 
data. QCD is the theory that describes {\sl quarks}, the 
smallest building blocks of matter according to the current theory,
and {\sl gluons}, the particles that quarks exchange.
Quarks are permanently bound together by a force due to the exchange of 
gluons. To test the QCD theory, large supercomputers have been built
to simulate what QCD will predict for physical parameters such as particle
masses. The predictions can then be compared against experiments to check
the reliability of the theory. The QCDSP computers at Columbia University 
and the RIKEN-BNL Research Center total 20,000 processors and are in 
full-time use for QCD simulations. 
The simulations are done on $4$D hyper-cube grid (discretization of 
space and time into a lattice). The $4$D data has to be 
analyzed for consistency and for correctness of the particle properties 
calculations.

\begin{figure}[ht]
\centerline{
    \psfig{figure=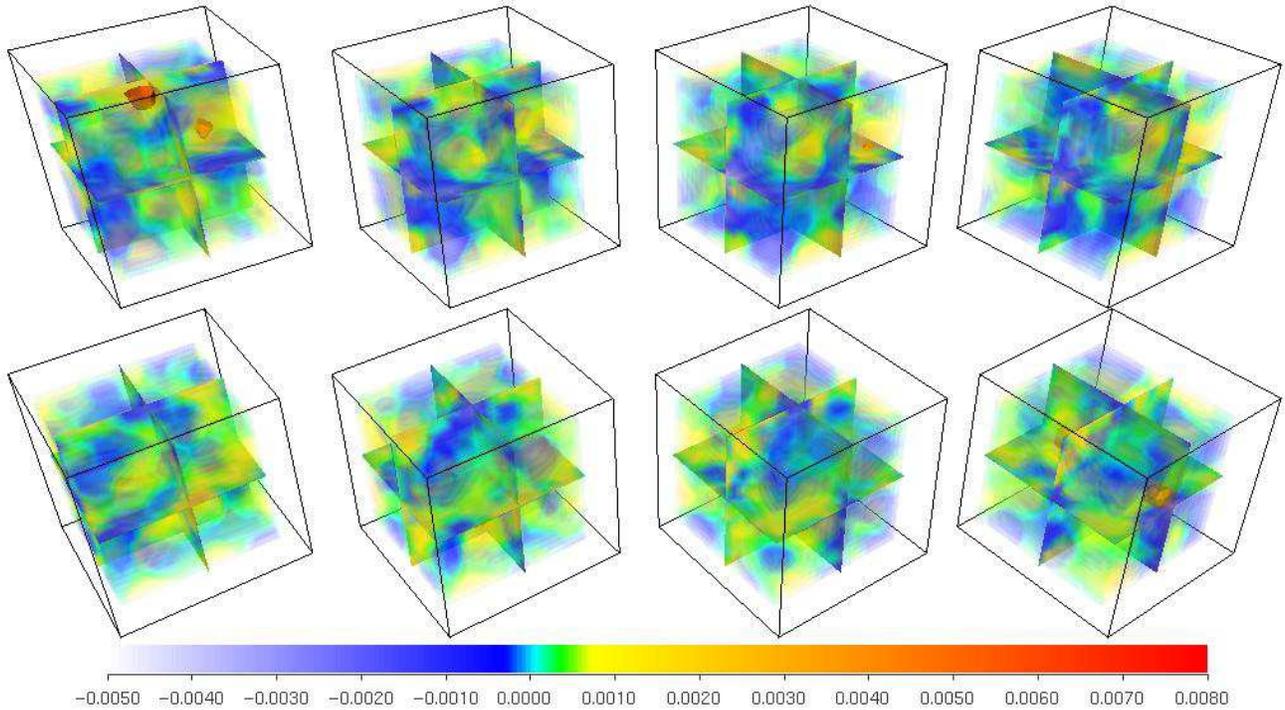,width=7.2in} 
           }
\caption{Multiview visualization of QCD data. Shown is topological charge 
density on a 4D $xyzt$ lattice of size 
$16$ X $16$ X $16$ X $16$. 
We show volume visualization of slices for $t = 1 .. 8$. The ordering is
from left to right, top to bottom. There is one iso-surface at $0.005$.}
\label{QCDVisualization}
\end{figure}

Another application is in material studies. In particular, we study 
meteorites' porosities, the presence and location of high 
attenuation cores, etc. To better understand the data we
use interactive synchronized views of the 3D meteorite data. 
For example, these are views of the volume visualization of
the meteorites' porous space, high-attenuation cores, volume
visualization of the entire data, cutting planes, and isosurfaces
(see Section \ref{results}, Figures \ref{meteo_cores} and \ref{meteo_pores}).

There are several approaches to visualize higher dimensional data. 
The most popular are parallel coordinates, slicing, and projections. 
Parallel Coordinates is a method of displaying
multivariate data \cite{inselberg}. Given an N-dimensional data set, 
N vertical axes are created. Each data point is represented as a polyline 
spanning all N axes. This is very useful for data sets of large dimension 
as several variables can be tracked at once. The slicing approach used in
\cite{asimov, wijk, hibbard} fixes one dimension or generates an animation 
to visit all values. The projection approach 
\cite{noll, andrews, hanson} projects all data points onto a
lower dimensional plane where their structure can be viewed and manipulated. 
User interfaces for  manipulating the projections are defined in 
\cite{duffin}. 

Our approach, as mentioned above, is to interactively visualize the full 
data, or at least as much as possible. For data in higher than 3 dimension, 
for example the QCD 4D data, we reduce the dimension by projecting or slicing
the data to 3D and displaying multiple views of the tightly coupled 
projections (or slices). The interactions are synchronized and yield
an easy and thorough view of the data. An internally defined language
allows the user to dynamically change view parameters and content.  
For example, Figure \ref{QCDVisualization} shows the tightly coupled 
visualization of 8 consecutive time slices of QCD data. We see a volume
visualization of the slices and, in this case, user added cutting planes,
bounding boxes, color bar, and isosurfaces. For 3D data, for example as
in the meteorites' study, an internally defined language
allows the user to dynamically add views and change their view 
parameters and content. The implementation is in C++. We use
the VTK \cite{vtk,vtkuserguide} API. Our graphical user interface
(GUI) uses VTK's interactors, interaction styles that
we developed, internally defined query language, and Tcl/Tk.
To handle many slices and still keep a high-resolution we
use a {\sl IBM T221} display, which maintains $3840$ X $2400$ pixels.

The rest of this paper is organized as follows. In Section \ref{model}
we give our multiview model and give some implementation details.
Section \ref{results} summarizes the visualization results and
the benefits from applying our multiview model. The results are explained 
within the QCD's and meteorite's applications.
Section \ref{extensions} discusses some of our current work on 
extensions relevant to multiview visualization and ideas for future work. 
Finally, in Section \ref{conclusions}, we give the conclusions of the 
studies presented.


\section{The multiview model and its implementation}\label{model}


We described briefly our multiview visualization model/approach in
the introduction (Section \ref{introduction}). In this section we 
present in more detail its main features:
\begin{itemize}
   \item ~Different interaction styles (Section \ref{styles}).
   \item ~Internally defined interaction language (Section \ref{language}).
   \item ~Multiple views (Section \ref{views}).
\end{itemize} 
We also give some implementation details in Section \ref{implementation}.

\subsection{Interaction styles}\label{styles}

We use the VTK's {\tt vtkInteractorStyle} class and its subclasses.
The subclasses provide diverse styles of mouse interaction: joystick/trackball
and camera/object. In camera mode for example, mouse events affect the camera
position and focal point, and hence all views in the current scene. 
In object mode, the user can pick an object and manipulate it
separately from the others in the scene. The interaction styles are changed 
with key strokes. Another useful feature is key binding that allows
the user to toggle the render window into and out of stereo mode. 
More information about the
interaction styles can be found in \cite{vtk,vtkuserguide} or in
online user manuals.

We implemented another subclass of {\tt vtkInteractorStyle}, called
{\tt vtkInteractorStyleActor}. In a standard VTK object mode interaction,
the mouse affects the object that is under the mouse pointer. Here
all the scene objects are affected, yielding a simultaneous
synchronized view of the objects from the same angle (see 
Section \ref{implementation}).

\subsection{Internal interaction language}\label{language}

We have internally defined an interaction language that allows
the user to dynamically change view parameters and view content.
The language controls view parameters such as object properties, 
color tables (including multiple color palettes), opacity scheme, background,
etc. The editing of the viewing content includes adding/removing
of views and objects in these views. The positions of the views are controlled
by the user. Objects that can be added/removed from a view are volumes, 
isosurfaces, cutting and viewing planes, colorbars, histograms, etc.
We have added to the language several filters that are specific to our 
data and visualization problems. Also, the language provides 
commands that facilitate the creation of user defined animations.

\subsection{Multiple views}\label{views}

Our model provides capability for multiple views. User's
needs, intent, and implemented interaction model 
determine and give control over the visualization content. 
The content may be obtained from the high dimensional data through 
(1) different user defined projections, see for example \cite{qcd},
(2) slicing of the high dimensional data, 
or (3) different views of the same 3D data. We provide 
``dimension reduction through projections'' by VTK's filtering 
capabilities, which are included in our internally defined interaction 
language. VTK's pipeline execution
model makes the projection implementation simple and efficient.
Slicing, which we use in the analysis of the QCD data, is automated
to the point where the user specifies a sequence of slices to be displayed.
Different views of the same 3D data are applied in our meteorite studies
(see Section \ref{results}).

\subsection{Implementation}\label{implementation}

As mentioned before, the implementation is in C++ and uses the VTK API.
Our {\tt vtkInteractorStyleActor} is a subclass of the VTK's
{\tt vtkInteractorStyle}. {\tt vtkInteractorStyle} has a pointer to
the current renderer, from which we get the collections of all the
actors and volumes in the scene. Instead of manipulating one object, 
as in the object style interaction, we make a loop and manipulate all 
the actors and volumes in the current scene.

The internal language interaction is implemented trough the
``u Keypress'' of the interaction style. Pressing ``u'' with active 
visualization window invokes a user defined function. The user 
can type single commands or pass scripts of the internally defined
interaction language.

Different views are implemented by simply applying proper
{\tt AddPosition} and {\tt SetOrigin} to the actors and volumes
that have to be visualized in the views. {\tt AddPosition} gives
an initial translation, in our case performed in the $xy$ plane,
of the new objects. {\tt SetOrigin} fixes the center for rotations
that may follow, in our case set to be the center of bounding box
for the new object.

\section{Visualization results}\label{results}
   
We visualized and analyzed two sets of QCD data.
The first one is of topological charge density. Figure \ref{QCDVisualization}
shows a multiview visualization of such data. The picture is for 
4D $xyzt$ lattice of size $16$ X $16$ X $16$ X $16$. Shown is
the volume visualization of slices for $t = 1 .. 8$. The ordering is
from left to right, top to bottom. Using the internally defined visualization 
language we have added $3$ cutting planes through the center of the volumes.
Also, there is one iso-surface marking density $0.005$.
In this case the visualization allowed us to easily make
consistency checks of the 4D QCD. The other set of data that we used
is for particle propagation. In particular, we visualized the
propagation of {\sl pion}, the lightest  bound state of two quarks.
The calculation of the pion propagator was performed on a 
$16$ X $16$ X $16$ X $32$ $xyzt$ lattice. The theoretical results
were visually analyzed and shown to have the correct  physical
behavior.
 
We also used the tool developed to visualize and analyze $32$ sets of
meteorite data. The data was produced using synchrotron-based computed 
microtomography (CMT) at the Brookhaven National Synchrotron Light Source.
This study approach, in contrast to daunting experimental efforts to 
analyze through grinding through the meteorites a few micrometers at a 
time \cite{bonte}, is much easier. It needs visualization to study
the presence and location of high attenuation cores and low attenuation
pores. Different attenuations are mapped and reveal the presence of
different metals in the meteorites. 
 
\begin{figure}[ht]
\centerline{
    \psfig{figure=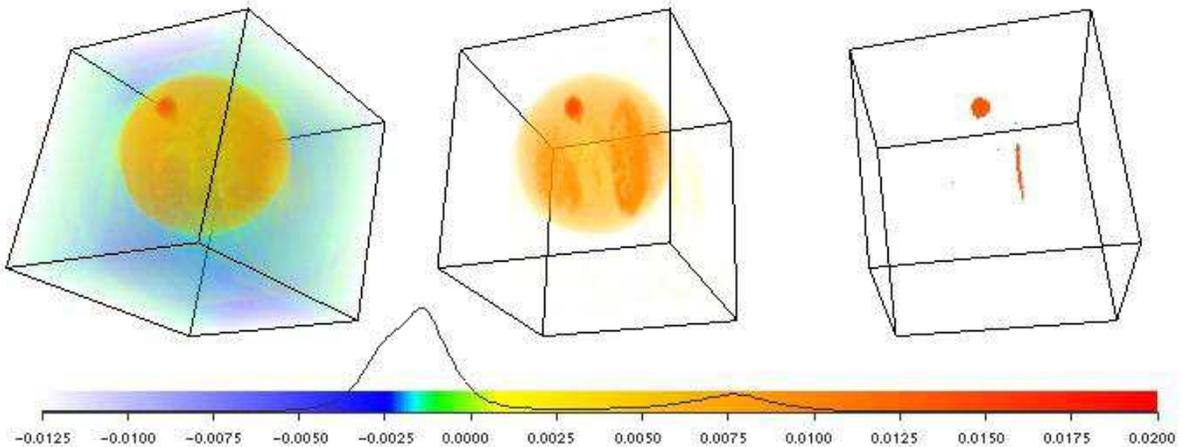,width=6.5in} 
           }
\caption{Multiview visualization of a meteorite's core.
         From left to right: (1) volume visualization of the whole data;
         (2) volume visualization of data range $0.002$..$0.02$; 
         (3) volume visualization of range $0.0125$..$0.02$.
         Used in studies for the presence and the relative location 
         of different metals that correspond to different attenuation 
         ranges of the data.}
\label{meteo_cores}
\end{figure}

Figure \ref{meteo_cores} shows $3$ views of a meteorite. 
From left to right we show: (1) volume visualization of the whole data;
(2) volume visualization of range $0.002$..$0.02$; and
(3) volume visualization of range $0.0125$..$0.02$. 
To determine the ranges we use the color bar and the histogram, which can  
also be seen on Figure \ref{meteo_cores}. The first peak is for the 
empty space around the meteorite. To skip it from the visualization and see 
only the core we visualize range $0.002$..$0.02$ in (2). One can see 
clustering of high-attenuation values within the meteorite's core.
Finally we show only the high-attenuation volume, which is after the 
second peak, i.e. range $0.0125$..$0.02$.

\begin{figure}[ht]
\centerline{
    \psfig{figure=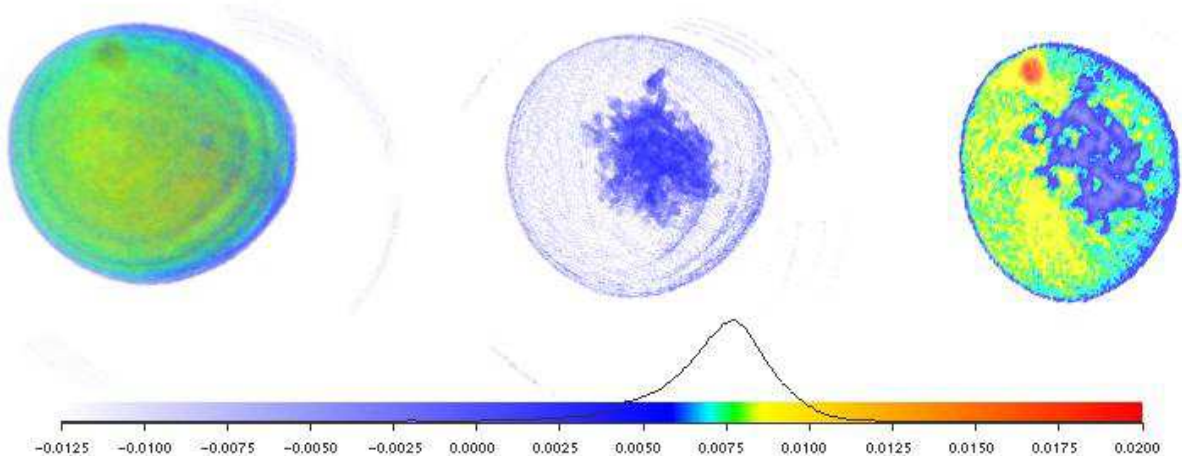,width=6.3in} 
           }
\caption{Multiview visualization of a meteorite's porous space.
         After filtering the data associated with air (see Figure
         \ref{meteo_cores}) we study the meteorite's porous space.
         From left to right: (1) volume visualization of the filtered data;
         (2) volume visualization of range $0.001$..$0.005$ (this should 
             be the porous space);
         (3) a slice through the center of the high-attenuation core 
             (see Figure \ref{meteo_cores}).}
\label{meteo_pores}
\end{figure}

It is also interesting to study the porous space. Figure \ref{meteo_pores}
shows the porous space for the meteorite from Figure \ref{meteo_cores}.
First we filtered out the attenuations less than $0.0025$ (as being
air). See that the histogram now does not have the air peak. We
have from left to 
right: (1) volume visualization of the filtered data; 
(2) volume visualization of range $0.001$..$0.005$ (this should 
   be the porous space); 
(3) a slice through the center of the 
   high-attenuation core which reveals its position relevant to the pores 
   and the meteorite's core.

\section{Extensions and future work}\label{extensions}

We are currently working on improving the speed of the multiview 
visualization environment presented. 
Adding new views increases the scene to be rendered 
and becomes a speed bottleneck for the interactive visualization.
We are looking into parallel methods to improve the speed and increase
the data size to be handled.
The task of multiview visualization is trivial to parallelize. 
We apply the idea from \cite{interactive} to extend this project to support 
parallel rendering on commodity-based clusters. We are interested in
both tiled displays and high-resolution flat panels.

\section{Conclusions}\label{conclusions}

We have developed a multiview interactive visualization for high dimensional
data. The functionality of the model was motivated by our need to study
QCD and meteorite data. The software and its functionality were tested 
on this data. As a result we were able to extract insight from the
data considered. In particular, the visualization allowed us to easily make
consistency checks of the 4D QCD data and to infer the correctness of
particle properties calculations. In the meteorite studies we inferred 
properties about the porous space and high-attenuation cores, which may
help in theories of how the meteorites were oxidized while entering the
Earth's atmosphere.

\vspace{0.1in}
{\bf {\large Acknowledgments}}.~
We would like to thank Robert Mawhinney, Cheng-Zhoung Sui,
Kostas Orginos, Shigemi Ohta, and Chris Dawson from
the RIKEN BNL Research center, and Keith Jones from the
BNL's Environmental Research and Technology Division
for the discussions with them and for providing the 
QCD and meteorites' data. The pion propagation calculation was
performed on the QCDSP RIKEN-BNL-Columbia Supercomputer, the meteorites'
data was produced by using synchrotron-based computed microtomography
at the Brookhaven National Synchrotron Light Source, and the visualization 
was developed at the BNL Visualization facility.               
\vspace{0.2in}


\end{document}